# Lost Audio Packets Steganography: The First Practical Evaluation

Wojciech Mazurczyk
Warsaw University of Technology, Institute of Telecommunications
Warsaw, Poland, 00-665, Nowowiejska 15/19

**Abstract.** This paper presents first experimental results for an IP telephony-based steganographic method called LACK (Lost Audio PaCKets steganography). This method utilizes the fact that in typical multimedia communication protocols like RTP (Real-Time Transport Protocol), excessively delayed packets are not used for the reconstruction of transmitted data at the receiver, i.e. these packets are considered useless and discarded. The results presented in this paper were obtained basing on a functional LACK prototype and show the method's impact on the quality of voice transmission. Achievable steganographic bandwidth for the different IP telephony codecs is also calculated.

## 1. Introduction

   VoIP (Voice over IP) is a real-time service which enables users to make phone calls through data networks that use an IP protocol. Generally, VoIP connection consists of two phases: a signalling phase and a conversation phase. In both phases certain types of traffic are exchanged between calling parties. After the signalling messages e.g. SIP (Session Initiation Protocol) [37] messages are exchanged between the caller and callee, and the connection is successful, the conversation takes place, in form of audio streams - RTP (Real-Time Transport Protocol) [2] streams - which are sent bidirectional. The popularity of this technology has caused a continuous rise in the volume of VoIP traffic. Thus, it may be increasingly targeted for steganographic purposes [35]. Steganographic methods allow for hiding the very existence of the communication, so a third-party observer will not suspect anything if they are unaware of the steganographic exchange. Steganography encompasses information hiding techniques that embed a secret message (steganogram) into the carrier.
   Lost Audio PaCKets steganography (LACK) is an IP telephony steganographic method, which modifies both: RTP packets from the voice stream, and their time dependencies. This method takes advantage of the fact that in typical multimedia communication protocols, like RTP, excessively delayed packets are not used for the reconstruction of transmitted data at the receiver, i.e. the packets are considered useless and discarded. LACK was originally proposed in [17] and studied further in [16]. The contribution of this paper is the practical evaluation of the influence that LACK has on voice transmission quality. Further advances involve the assessment of its potential steganographic bandwidth for different IP telephony codecs. This was achieved by means of constructing a LACK prototype and conducting appropriate experiments at different levels of intentional losses.
   The detailed overview of LACK functioning is presented in Fig. 1. At the transmitter (Alice), one RTP packet is selected from the voice stream and its payload is substituted with bits of the secret message – the steganogram (1). Then, the selected audio packet is intentionally delayed prior to its transmission (2). Whenever an excessively delayed packet reaches a receiver unaware of the steganographic procedure, it is discarded, because it interprets the hidden data as "invisible". However, if the receiver (Bob) is aware of the hidden communication, then, instead of dropping the received RTP packet, it extracts the payload (3). Due to the fact that the payload of the intentionally delayed packets is sole vector used to transmit secret information to receivers aware of the procedure, therefore no surplus packets are generated.

LACK is a TCP/IP application layer steganography technique and is fairly easy to implement. This may be attributed to the fact that RTP is usually integrated into telephone endpoints (softphones), therefore access generation and modification of RTP packets is easier to perform than in the case of the lower layer protocols like IP or UDP.

LACK, as any network steganography method, can be characterised by the following set of features: its steganographic bandwidth, undetectability and its steganographic cost. The term - steganographic bandwidth - refers to the amount of secret data that we are able to send per time unit when using a particular method. Undetectability is defined as the inability to detect a steganogram within a certain carrier. The most popular way to detect a steganogram is to analyse statistical properties of the captured data and compare it with the typical properties of that carrier. The steganographic cost characterises the degree of degradation of the carrier caused by the steganogram insertion procedure. For LACK, this cost can be expressed by means of providing a measure of conversation degradation.

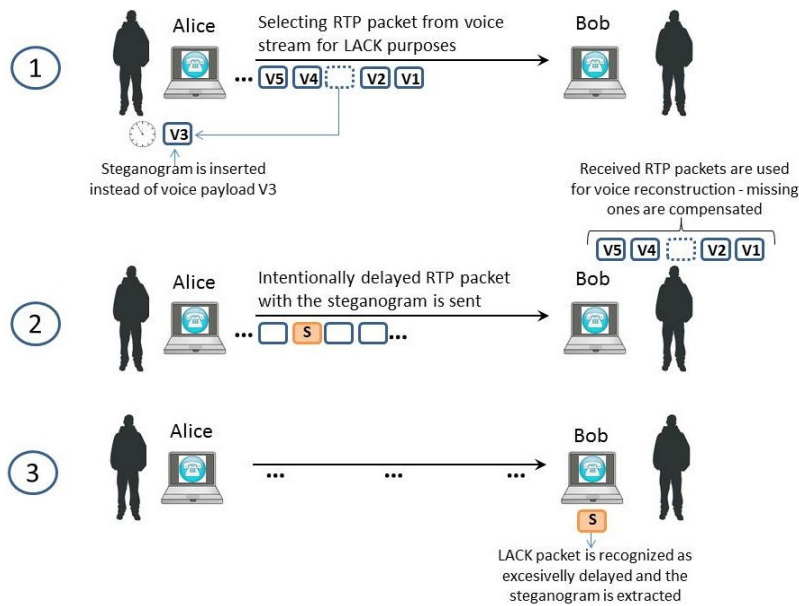

Fig. 1 The idea of LACK

Steganalysis of LACK is hard to perform because packet loss in IP networks is a "natural phenomenon". A packet is considered lost if: it is discarded in the network– in this case it never reaches the receiver. Such a situation may be caused, e.g., by buffer overflow in some intermediate device caused by a bottleneck within a network; it is dropped by the jitter buffer– when an RTP packet is excessively delayed due to network latency it reaches the receiver but is useless as it cannot be used for voice reconstruction; thus, it is discarded and counted as lost. Moreover, due to so-called delay spikes, the jitter buffer, in addition to dropping late packets (drops caused by buffer underflow) may also drop subsequent RTP packets because they may all arrive simultaneously and the size of the jitter buffer may be insufficient to store them all (buffer overflow). Results from study in [38] revealed that about 80% of performed Internet calls experienced about 0.5% of physical RTP packet losses and about 30% of the call 2% or more jitter buffer losses. Therefore, intentional losses introduced by LACK are not easy to detect, if kept on a reasonable level. Potential LACK steganalysis methods include:

- Statistical analysis of lost packets for calls in a sub-network. This type of steganalysis may be implemented with a passive warden [18] (or some other network node), based, for example, on information included in RTCP reports (the cumulative number of packets lost field) exchanged between users during their communication or by observing RTP stream flows (packets' sequence numbers). If for some of the observed calls the number of lost packets is higher than average (or some chosen threshold), this criterion may be used as an indication for the potential use of LACK.
- Statistical analysis based on the VoIP calls duration. If the call duration probability distribution for a certain sub-network is known, then statistical steganalysis may be performed to discover VoIP sources that do not fit to the distribution (the duration of LACK calls may be longer compared to non-LACK calls as a result of introducing steganographic data).
- An active warden [18] that analyses all RTP streams in the network (SSRC identifier and fields: Sequence Number and Timestamp from RTP header) can identify packets that are already too late to be used for voice reconstruction. The active warden may erase their payload fields or simply drop them. A potential problem that arises in this case is to avoid eliminating delayed packets that may still be used for conversation reconstruction. The size of the jitter buffer at the receiver is, in principle, unknown to the active warden. If an active warden drops all delayed packets, then it will potentially drop packets that still can be useful for voice reconstruction. In effect, the quality of conversation may deteriorate considerably. Moreover, not only steganographic calls are affected; non-steganographic calls are also "punished".

If the VoIP call is secured using SRTP [39] it has no influence on LACK utilisation. Moreover, it makes LACK even less susceptible to detection. It is due to fact that even if warden captures all RTP packets it will not be able to reconstruct voice conversation because it is encrypted and thus it will be unable to spot steganogram inside these packets.

The rest of this paper is structured as follows. Section 2 presents existing VoIP steganographic methods. Section 3 discusses the factors that have impact on the LACK steganographic bandwidth, its undetectability and cost. Section 4 describes the prototype LACK implementation, experiment methodology and discusses the obtained results. Section 5 concludes this work.

## 2. Related Work

IP telephony as a hidden data carrier was discovered by researchers rather late. Proposed steganographic methods have been developed, generally, from two distinctive research origins. Firstly, from the well-established image and audio files steganography [19] – these methods targeted voice digital representation as a hidden data carrier. Secondly, from the covert channels created in different network protocols [20, 21] – these solutions targeted specific VoIP protocols fields (e.g. signalling protocol – SIP, transport protocol – RTP or control protocol – RTCP) or the way these protocols behave and interact. Today, steganographic methods that can be used in telecommunication networks have been described by term *network steganography* or when applied, specifically, to IP telephony by term *steganophony* [35].

First VoIP steganographic methods that utilised carried voice as a hidden data carrier were proposed by Dittmann et al. in 2005 [22]. Authors proposed evaluation of existing audio steganography with a special focus on solutions which are suitable for VoIP. This work was later extended and published in 2006 in [23]. In [25] implementation of the SteganRTP tool was described which to carry steganograms used least significant bits (LSB) of G.711 codec. Wang and Wu in [26] also suggested using least significant bits in voice samples but to carry bits of secret communication which was coded using lower rate voice codec like Speex. In [27] Takahashi and Lee introduced similar approach by presenting proof of concept tool – Voice over VoIP (Vo2IP) that can establish a hidden conversation by embedding further compressed voice data into regular PCM-based voice traffic. They also considered other audio steganography methods that can be utilized in VoIP like DSSS, FHSS or Echo hiding. Aoki in [28] proposed steganographic method based on the characteristics of PCMU in which 0 speech sample can be represented by two codes due to the overlap. Another LSB-based method was proposed by Tian et al. in [32]. Authors incorporated the m-sequence technique to eliminate the correlation among secret messages to resist the statistical detection. Another similar approach (also LSB-based) for adaptive VoIP steganography by the same authors was introduced in [33]; a proof of concept tool - StegTalk – was also developed. In [34] Miao and Huang presented an adaptive steganography scheme that is based on smoothness of the speech block. Such approach proved to give better results in terms of voice quality than LSB method.

Utilisation of the VoIP-specific protocols as a steganogram carrier was first proposed by Mazurczyk and Kotulski in [24] to embed control information into the VoIP streams. Unused bits in the headers of IP, UDP and RTP protocols carry the type of parameters and actual parameter values are embedded as watermark in the voice data. In [29] and [17] Mazurczyk and Szczypiorski described steganographic methods that can be used for VoIP signalling protocol – SIP (with SDP) and RTP streams (with RTCP), respectively. They discovered that when combined steganographic methods during signalling phase are able to transfer about 2000 bits of steganogram and during the conversation phase about 2.5 kbit/s. Bai et al. in [31] proposed covert channel based on jitter field of the RTCP header. First, statistical parameters of the jitter field in the current network are calculated. Then, the secret message is modulated into the jitter field according to the previously calculated parameters. By utilising such modulation the characteristic of the covert channel is similar to that of the overt one. In [36] Forbes proposed new RTP-based steganographic method that modifies timestamp value in the RTP header to send steganograms. The method steganographic bandwidth is theoretically up to 350 bit/s.

## 3. Factors that Impact LACK Performance

The general principle in LACK is that the more hidden information is inserted into the voice stream, the greater the chance that it will be detected, i.e. by scanning the data flow or applying some other steganalysis (detection) method. Secondly, the more audio packets are used to send covert data, the greater the deterioration of the quality of IP telephony connection. This, in turn, results in a greater steganographic cost. Therefore, the procedure of the insertion of hidden data must be carefully chosen and controlled in order to minimize the chance of the detection of the inserted data and to avoid excessive deterioration of the QoS (Quality of Service). That is why certain trade-offs between the achieved steganographic bandwidth, call quality deterioration and resistance to detection are (always) indispensable.

The performance of LACK depends on many factors which can be divided into three following groups:
- *Endpoint-related* factors: the type of voice codec used (in particular, its resistance to packet losses and the default voice quality), size of the RTP packet payload and the size of the jitter buffer.
- *Network-related* factors: packet delay, packet loss probability and jitter.
- *LACK-related* factors: the number of intentionally delayed RTP packets, the delay of the LACK packets and hidden data insertion rate (*IR*), which corresponds to the number of steganogram's bits carried per unit of time [bit/s].

**3.1 Endpoint-related factors**

To guarantee that an RTP packet will be deemed lost by the receiver, it must be excessively, intentionally delayed by the LACK procedure. To set this delay $d_L(t)$ properly, the size of the receiver's jitter buffer must be taken into account. A jitter buffer is used to alleviate the jitter effect, i.e. the variations in packets arrival time caused by queuing, contention and serialization in the network. The size of the buffer is implementation-dependent. It may be fixed or adaptive, and is usually between 60 and 120 ms. An RTP packet will be recognized as lost whenever its delay exceeds the delay introduced by the jitter buffer. LACK users must exchange information about the sizes of their jitter buffers prior to starting the hidden communication. To limit the risk of disclosure of a steganogram, the delay chosen by LACK should be as low as possible. The delay of an RTP packet ($d_T$) may be calculated as follows

$$d_T(t) = d_D + d_K + d_E + d_L(t) \qquad (2\text{-}1)$$

where:
$d_D$ – delay introduced by DSP (Digital Signal Processor), which depends on the type of the codec and typically ranges from 2 to 20 ms,
$d_K$ – delay introduced by voice coding (typically under 10 ms),
$d_E$ – delay caused by encapsulation (from 20 to 30 ms).
$d_L(t)$ – the intentional delay of an RTP packet introduced by LACK,

As mentioned above, the value of the intentional delay $d_L(t)$, introduced by LACK, must be carefully chosen. Together with $d_N(t)$, introduced by the network, it must exceed the size of the jitter buffer (Fig. 2), that is

$$d_T(t) + d_N(t) > t_B(t) \qquad (2\text{-}2)$$

where:
$d_N(t)$ – delay introduced by the network,
$t_B(t)$ – the size of the jitter buffer

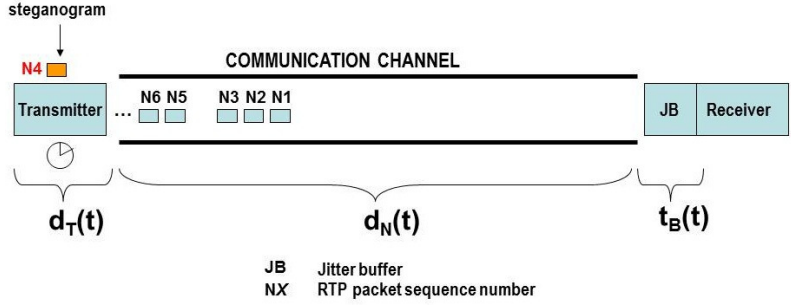

Fig. 2 The LACK delay components

The jitter buffer can be of a fixed or adaptive size. If the jitter buffer has a fixed size during the call, and it does not take into account information about delay caused by network then delay at the transmitter side should be

$$d_T \geq t_B \qquad (2\text{-}3)$$

and

$$d_L \geq t_B - d_D - d_K - d_E \qquad (2\text{-}4)$$

Consequently, a fixed size jitter buffer that reacts to the current delay $d_N(t)$, introduced by the network during the call, then the delay at the transmitter output is

$$d_T(t) \geq t_B - d_N(t) \qquad (2\text{-}5)$$

thus

$$d_L(t) \geq t_B - d_D - d_K - d_E - d_N(t) \qquad (2\text{-}6)$$

If the current value of $d_N(t)$ is not known at the transmitter, then one can utilise the average value of the delay calculated over a certain time period.

If the adaptive jitter buffer is used at the receiver, and the information regarding its size is not passed to the transmitter during the call, then the relation $d_T(t) \geq t_B(t)$ should be fulfilled. This is to ensure that intentionally delayed RTP packets will not be used for voice reconstruction. Due to the fact that delays $d_D$, $d_K$ and $d_E$ are constant, then ensuring $d_T(t) \geq t_B(t)$ is possible when

$$d_L(t) \geq t_B^* - d_D - d_K - d_E \qquad (2\text{-}7)$$

where $t_B^*$ denotes the maximum, admissible size of the adaptive jitter buffer.

Under the considered conditions, if the receiver is equipped in an adaptive jitter buffer and it is possible to advertise its size during the call, then its initial size can be communicated during the signalling phase of the call. This imposes that the delay at the transmitter output $d_T(0)$ is set equal to the maximum possible size of the jitter buffer – $d_T(0) \geq t_B^*$. It can be further decreased, by means of reducing $d_L(t)$, which is possible if appropriate information about the variations in size of the jitter buffer reaches the transmitter during the call.

When an adaptive jitter buffer is employed, and the transmitter is informed of the current network delay $d_N(t)$ then

$$d_L(t) \geq t_B(t) - d_D - d_K - d_E - d_N(t) \tag{2-8}$$

The other factor that influences LACK is the VoIP codec used for the conversation. The greater codec resistance to packet losses the more favourable it is for LACK purposes. The admissible level of packet losses for different voice codecs usually ranges from 1 to 5%. For example, according to [12], the maximum loss tolerance equals 1% for G.723.1, 2% for G.729A and 3% for G.711 codecs. The usage of mechanisms that deal with lost packets at the receiver, e.g. the PLC (Packet Loss Concealment) [3] results in an increase in the acceptable level of packet losses, e.g. for G.711 the shift is from 3% to 5%. The greater the codec resistance to packet losses, the greater the capacity for achieving a significant LACK steganographic bandwidth. Thus, the quantity of covert data liable for insertion by LACK procedure and, consequentially, the additional, induced packet losses depends on the acceptable level of the cumulative packet loss.

It is also worth noting that the use of silence suppression mechanism in the transmitting endpoint can further decrease the available steganographic bandwidth in which to hide secret messages.

**3.2 Network-related factors**

Let us assume that, at a given moment of the call $t$, an RTP packet is chosen from the voice packets stream for LACK purposes with probability $p_L(t)$ and the network packet loss probability is $p_N(t)$. If $p_T$ denotes the maximum permitted probability of RTP packet losses then, assuming the independence of the network-related losses from LACK-induced losses, we get

$$p_T \leq 1 - (1 - p_N(t))(1 - p_L(t)) \tag{2-9}$$

and, in consequence

$$p_L(t) \leq \frac{p_T - p_N(t)}{1 - p_N(t)} \tag{2-10}$$

which describes the admissible level of RTP packet losses introduced by LACK. Exemplary relationships between probabilities $p_L(t)$, $p_N(t)$ and $p_T$ are illustrated in Fig. 3.

To ensure a high steganographic bandwidth and the undetectability of LACK, it is necessary to monitor network conditions while the call lasts. In particular, packets losses, delay and jitter introduced by the network must be observed. They have influence on the range of permitted values of the delay and packet losses introduced by LACK without the degradation of the perceived quality of the conversation. Due to the fact that LACK exploits for its purposes legitimate RTP traffic, an increase in the overall packets losses is triggered. Thus, the number of lost packets used for steganographic purposes must be controlled and dynamically adapted.

Information concerning current network conditions can be provided to the transmitter, among others, with the aid of SR (Sender Report), RR (Receiver Report) [2] or XR (Extended Report) [14] reports that are defined in the RTCP protocol. The lack of monitoring of network parameters during a call does not hinder the possibility of determining their values, what can be achieved with the aid of historical, statistical data related to the network performance. However, it should be noted that RTP packet losses introduced by network can provoke a decrease in the LACK steganographic bandwidth, which is the case if the lost packet belongs to the steganographic RTP stream.

**3.3 LACK-related factors**

In previous subsections we mentioned two important parameters that require setting for proper LACK functioning. These are: the probability that a certain RTP packet is chosen for LACK purposes ($p_L(t)$), and the delay $d_L(t)$, which is preset to guarantee that an audio packet will be recognized as lost at the receiver. Another key factor influencing LACK steganographic bandwidth and its resistance to steganalysis is the hidden data insertion rate *IR(t)*, which is defined as a number of steganograms' bits carried in every time unit of the call [bit/s]. In general, the greater *IR(t)*, the greater is the achievable steganographic bandwidth. This couples with the degradation of voice quality and easier steganalysis. The limits imposed on the maximum insertion rate depend on the targeted acceptable call quality, network conditions, the size of the steganogram and also by the duration of the call. The correct determination of *IR(t)* facilitates efficient control of RTP packet losses and delays introduced by LACK, without excessively deteriorating the call quality and risking detection. The methods for determining *IR(t)* based on current conversation quality, the size of the steganogram and the duration of the call were considered in [16].

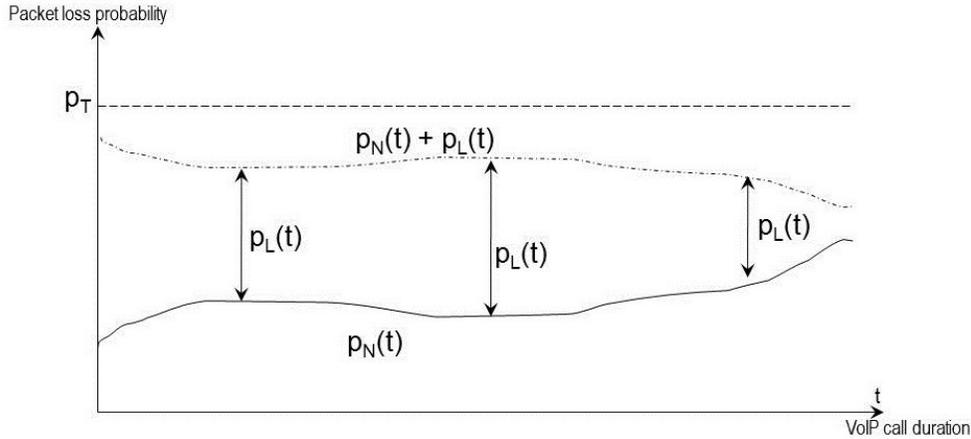

Fig. 3 The impact of LACK on the total packet loss probability

## 4. Practical Evaluation of LACK's Impact on Call Quality

Call quality may be expressed in terms of subjective and objective quality measures. Objective measures are usually based on algorithms such as the E-Model [4], PAMS or PESQ [6]. The objective measures can be transformed into subjective quality measures. In our analysis we shall use the subjective measure MOS (Mean Opinion Score) [5] which is calculated by PESQ method.

**4.1 LACK Prototype and Experiment Methodology**

Implementation of LACK prototype was based on the *MjSip* [9] project. It is a Java implementation of an IP telephony softphone based on the SIP (Session Initiation Protocol) signalling protocol. Only the user agent application was utilised. The SIP server was omitted because it does not affect the results of experiments (the RTP streams are exchanged directly between end users, without the participation of a SIP server). A simple PLC (Packet Loss Concealment) method was implemented in the SIP User Agent application, as softphones usually have some way to deal with packets losses. PLC mechanisms are used to limit quality degradation caused by packet losses by means of compensating the lost ones – in the simplest scenario, they insert a duplicate of the last received packet as a substitute for the lost one [3]. The described PLC algorithm was added to the SIP User Agent application.

The LACK algorithm proved to be easily implementable, and its main principles shall be described in the following clause. In our implementation, every RTP packet selected for LACK purposes had a payload consisting of two parts: a steganogram and a hash. The hash was computed for the steganogram carried in that packet with the aid of the MD5 (Message Digest 5) hash function. The amendment of the hash enables the receiver to distinguish LACK packets from ordinary, non-steganographic packets.

Two parameters of the LACK method were studied: the probability that a packet is used for LACK purposes ($p_L$) and the intentional delay of LACK packets ($d_L$). For each RTP packet, a pseudo-random number between 0 and 1 was generated, and it was tested whether it exceeds the threshold probability of sending a LACK packet. If this was the case, the considered packet was chosen for steganographic purposes.

During the deployment of LACK, a problem concerning the inaccuracy of the sleep function in Java was encountered. This function was used to determine the time interval between consecutive packets. The inaccuracy of the Java function resulted in imprecise timings of data packets. We solved the problem by introducing compensation of the delays at the receiver.

**4.2 Experiment Methodology**

The experimental setup used to evaluate LACK's impact on voice quality is presented in Fig. 4. The experimental environment was a controlled LAN network, so no RTP packets were lost or excessively delayed, unless intended. No RTP packets were dropped by the jitter buffer, which permitted us to evaluate the sole impact of LACK on voice quality, without any network-related or endpoint-related interferences.

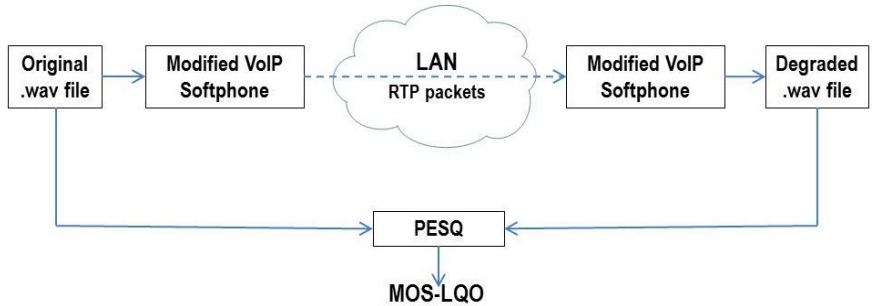

Fig. 4 MLS experimental setup

The voice packet payload was selected from TIMIT [8] speech samples database and compiled into single .wav file. Both male and female voices were used. The resulting .wav input file was about 30 seconds long. The adopted coding involved PCM, 8000 Hz sampling, 16 bit sample representation of monophonic signal. It was ensured that the setup conformed to ITU-T P.862.3 recommendation [7] requirements fulfilling which ensure proper PESQ method functioning. The obtained results were normalized to 9 minutes, as it was experimentally verified that the average call duration for IP telephony falls in the range 7-11 minutes [11]. The 9 minute representation was chosen to show how much secret data can be sent during a typical IP telephony call. The input .wav file was encoded with different voice codecs and its parts were then inserted into the payloads of consecutive RTP packets. Table I provides a summary and characteristics of the voice codecs used in the experiment. A variety of codecs were chosen to provide a comparative analysis of possible IP telephony call configurations – the choice involved selection of voice codec and appropriate data rate (from 8 to 64 kbits). The voice codecs used in the experiment were: G.711 A-law [3], GSM-FR (Full Rate) [1] and Speex (8 and 24.6 kbits) [13].

Table I Speech codecs used in the experiment

| Voice codecs | G.711 A-law | Speex I | GSM-FR | Speex II |
|---|---|---|---|---|
| Bit rate [kbit/s] | 64 | 24.6 | 13.2 | 8 |
| RTP packet every [ms] | 20 | 20 | 20 | 20 |
| Voice payload size [bytes] | 160 | 61.5 | 33 | 20 |

The second step of the procedure involved the modification of the RTP stream by means of performing the LACK steganographic method (the introduction of intentional losses, for which the selection probability, $p_{LACK}$, was picked from the range 0.001 to 0.05). Next, the RTP stream was directed to the receiver, where the voice conversation was reconstructed and saved to the output .wav file. Then, the original (input) and degraded (output) .wav files were compared with the aid of the PESQ method, and the MOS-LQO (Mean Opinion Score-Listening Quality Objective) value was obtained. By performing the experiments in a strictly controlled environment with no network or jitter buffer losses and without excessive delays, we were able to assess the real influence of LACK on the conversation quality. Each experiment was repeated 10 times, and the average results are presented in the following section.

### 4.3 Experimental Results

The obtained experimental results are presented in Table II and Fig. 5 and 6 (σ denotes standard deviation).

Table II Experimental results

| $p_{LACK}$ | G.711 A-law | | | | Speex I (24.6 kbps) | | | | GSM-FR | | | | Speex II (8kbps) | | | |
|---|---|---|---|---|---|---|---|---|---|---|---|---|---|---|---|---|
| | MOS-LQO | | $S_B$ [bit/s] | | MOS-LQO | | $S_B$ [bit/s] | | MOS-LQO | | $S_B$ [bit/s] | | MOS-LQO | | $S_B$ [bit/s] | |
| | Av. | σ | Av. | σ | Av. | σ | Av. | σ | Av. | σ | Av. | σ | Av. | σ | Av. | σ |
| 0.001 | 4.015 | 0.02 | 84.48 | 32.13 | 3.992 | 0.15 | 22.29 | 16.14 | 3.269 | 0.26 | 3.62 | 2.03 | 3.527 | 0.03 | 3.02 | 2.07 |
| 0.005 | 3.91 | 0.06 | 376.32 | 154.56 | 3.553 | 0.39 | 72 | 27.36 | 2.347 | 0.23 | 52.13 | 13.09 | 3.487 | 0.07 | 8 | 1.38 |
| 0.01 | 3.865 | 0.09 | 591.36 | 168.7 | 3.261 | 0.44 | 228 | 51.61 | 2.078 | 0.14 | 59.84 | 11.73 | 3.419 | 0.05 | 16.18 | 2.97 |
| 0.02 | 3.622 | 0.05 | 1236.48 | 235.78 | 3.059 | 0.21 | 376.8 | 18.2 | 1.884 | 0.18 | 134.19 | 28.78 | 3.28 | 0.07 | 39.73 | 8.05 |
| 0.03 | 3.597 | 0.11 | 1735.68 | 170.44 | 2.946 | 0.35 | 602.4 | 104.41 | 1.568 | 0.13 | 227.57 | 10.82 | 3.144 | 0.08 | 57.17 | 7.57 |
| 0.05 | 3.338 | 0.09 | 2872.32 | 178.88 | 2.876 | 0.59 | 1039.2 | 120.18 | 1.286 | 0.04 | 369.92 | 36.35 | 2.964 | 0.11 | 89.39 | 10.35 |

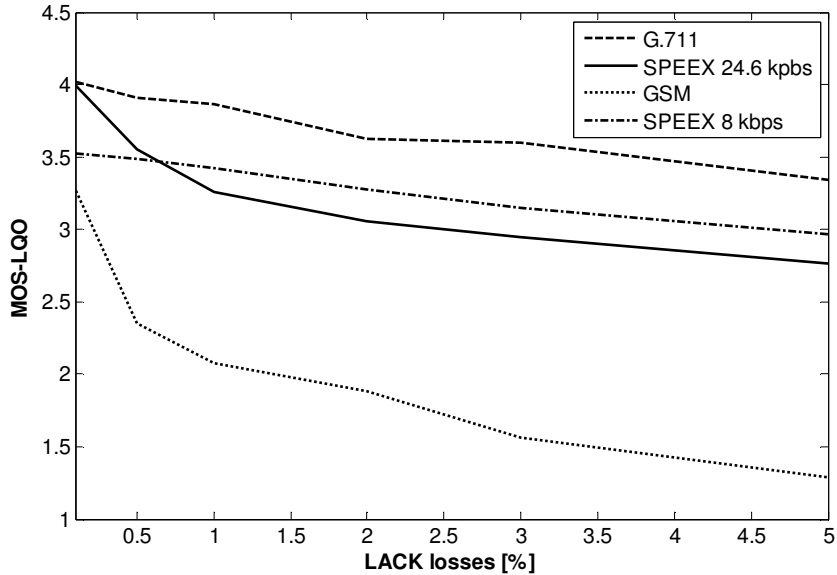

Fig. 5 Voice quality results (MOS-LQO)

As anticipated, the presented results prove that the best choice for LACK purposes is the G.711 codec, as it can sustain RTP packet losses exceeding 5% and still provide voice quality with a MOS score greater than 3 (which is considered as fair). Simultaneously, the G.711-based LACK provides the largest steganographic bandwidth, e.g. at the level of 1% of LACK packet losses it offers about 590 bit/s (Fig. 6). Such performance is achievable because the voice payload size in each RTP packet is 160 bytes, which is considerably more than for any other chosen codec (Speex I – 61.5, GSM – 33 and Speex II – 20 bytes).

Speex I codec with a bit rate of 24.6 kbit/s turned out to give worse results in voice quality (Fig. 5) than G.711 (64 kb/s). The surprising observation is that Speex I performance was worse than Speex's II (8 kbit/s) – when LACK loss rate exceeded 0.75% it experienced a 0.2 drop in MOS score relative to Speex II. On the other hand, tests performed for a 1% LACK loss rate proved that Speex I is capable of achieving a larger steganographic bandwidth (230 bit/s compared to 16 bit/s for Speex II), which may be attributed to a higher bitrate and larger payload size of this codec. Such performance is possible with the simultaneous preservation of a still higher than fair quality (MOS scores for Speex I – 3.26, Speex II – 3.42). The GSM FR codec offered the poorest voice quality even for low levels of introduced LACK losses (>0.1%) and thus it is least suitable for LACK purposes.

In general, the highest values of the steganographic bandwidth are achieved when LACK intentionally delays numerous RTP packets. Thus, the high bit rate codecs are preferred over low bit rate ones. Currently, one of the most popular voice codecs used almost in all IP telephony soft- and hard-phones is G.711, which proved to be suitable for LACK.

However, in real-life IP networks, LACK may not introduce as many intentional losses as in the experiment (i.e. significantly less than 5%). Causing excess losses can have a great impact on voice quality because the packet drops cumulate with the network and jitter buffer losses (see Section 3).

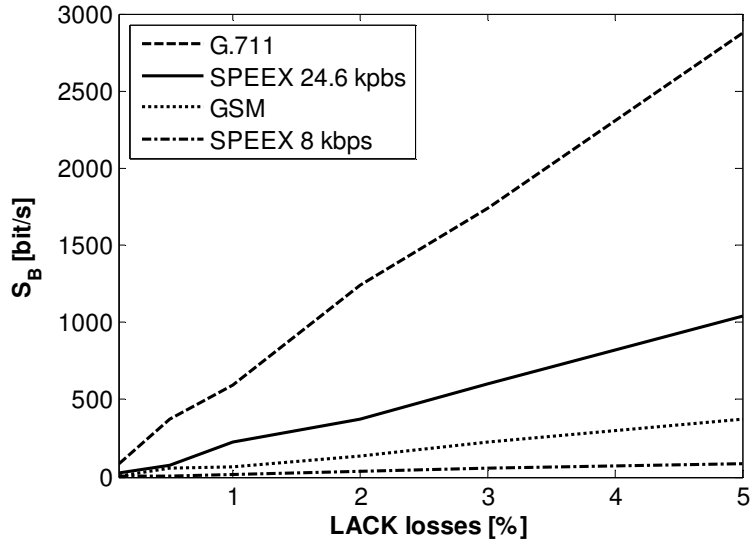

Fig. 6 LACK steganographic bandwidth

On the other hand, from the network security point of view, the usage of low bit rate voice codecs in IP telephony services can limit LACK steganographic bandwidth. For example, at a LACK loss rate equal to 0.1%, G.711 provided 85 bit/s steganographic bandwidth, while Speex II offered only about 3 bit/s. However, even such low value of steganographic bandwidth is still considered insecure [10], so additional steganalysis methods should be utilised to detect and/or eliminate hidden communication, e.g. methods that were proposed in [16].

## 5. Conclusions and Future Work

In this paper, an IP telephony steganographic method, named LACK, was subjected to the first practical evaluation. The aim of this paper was to study LACK's impact on the quality of voice transmission and the achievable steganographic bandwidth for different IP telephony codecs. This was considered in a broader context of the characterisation of factors that influence LACK steganographic bandwidth, cost and undetectability.

The obtained results show that from the LACK perspective the most favourable voice codecs are those of high bit rates (e.g. G.711). They are capable of accommodating higher steganographic bandwidths than low bit rate codecs, while being more immune to packet losses. This imposes that for such codecs the steganographic cost of utilizing LACK is lower. From the network security point of view, the deployment of low bit rate codes in IP telephony systems is preferable because it constrains the potential LACK bandwidth. However, low bit rate voice codecs cannot be treated as a universal solution for this type of covert communication. Still, to provide ability to detect or eliminate LACK hidden communication proper steganalysis method must be utilized.

Future work should involve an experimental evaluation of the LACK method in a real-life network environments e.g. in a WLAN (Wireless Local Area Network), with the employment of a wide range of voice codecs and different PLC mechanisms.

**ACKNOWLEDGMENTS**

- This work was supported by the Polish Ministry of Science and Higher Education under Grants: N517 071637 and IP2010 025470.
- The authors would like to thank:
    - Dr. Artur Janicki from Warsaw University of Technology (Poland) for help, valuable comments and fruitful discussions.
    - Wojciech Frączek from Warsaw University of Technology (Poland) for sharing his implementation of the LACK steganographic method.
    - Dr. Krzysztof Szczypiorski and Elżbieta Zielińska from Warsaw University of Technology (Poland) for helpful comments and remarks.